\documentclass[12pt]{iopart}

\usepackage{bbold}
\begin{document}

\title[Bayesian derivation of plasma EDF]{Bayesian derivation of plasma equilibrium distribution function for tokamak scenarios and the associated Landau collision operator}

\author{C. Di Troia}

\address{ENEA Unit\`{a} tecnica Fusione, C.R. Frascati, Via E. Fermi 45, 00044 Frascati (Rome), Italy}
\ead{ claudio.ditroia@enea.it}
\vspace{10pt}
\begin{indented}
\item18 January 2015
\end{indented}

\begin{abstract}
A class of parametric distribution functions has been proposed in  [{\sc C.~DiTroia}, 
 {\em Plasma Physics and Controlled Fusion}, {\bf 54}, (2012)] as equilibrium distribution functions (EDFs) for charged particles in fusion plasmas, representing supra-thermal particles in anisotropic equilibria for Neutral Beam Injection, Ion Cyclotron  Heating scenarios. Moreover, the EDFs can also represent nearly isotropic equilibria for Slowing-Down $alpha$ particles and core thermal plasma populations.
These EDFs  depend  on constants of motion (COMs). Assuming an axisymmetric system with no equilibrium electric field, the EDF depends on the toroidal canonical momentum $\mathcal{P}_\phi$, the kinetic energy $w$ and the magnetic moment $\mu$. \\
In the present work, the EDFs are obtained from first principles and general hypothesis. The derivation is probabilistic and makes use of the Bayes' Theorem.
 The bayesian argument allows us to describe how far from the prior \emph{probability distribution function} (pdf), \emph{e.g. Maxwellian}, the plasma is, based on the information obtained from magnetic moment and GC velocity pdf.\\
 Once the general functional form of the EDF has been settled, it is shown how to associate a \emph{Landau} collision operator and a \emph{Fokker-Planck} equation that ensures the system relaxation towards the proposed EDF.
\end{abstract}

% Uncomment for PACS numbers
%
% Uncomment for keywords
\vspace{2pc}
\noindent{\it Keywords}: tokamak, distribution function, Fokker-Planck, Bayes' theorem 
%
% Uncomment for Submitted to journal title message
%\submitto{\JPA}
%
% Uncomment if a separate title page is required
%\maketitle
% 
% For two-column output uncomment the next line and choose [10pt] rather than [12pt] in the \documentclass declaration
%\ioptwocol
%

\section{Introduction}

  The following parametric distribution function has been proposed in Ref.\cite{me} as equilibrium distribution function (EDF) for charged particles in fusion plasmas, representing, \emph{e.g.}, supra-thermal particle distribution produced by additional external heating sources in tokamak experiments: 
 \begin{equation}
 \label{feq}
f_{eq}=\frac{\mathcal{N}\left( w/T_w\right)^{\alpha_w}}{\sqrt{2\pi} w^{3/2}} \exp \left[-\left(\frac{\mathcal{P}_{\phi}-\mathcal{P}_{\phi0}}{\Delta_{P_\phi}}\right)^2 \right] \exp \left\{-\frac{w}{T_w}\left[1+\left(\frac{\lambda-\lambda_0}{\Delta_\lambda}\right)^2\right] \right\},
\end{equation}
 being  $w$ the kinetic energy per unit mass, $\mu$ the magnetic moment per unit mass, $\lambda=\mu/w$  the pitch angle and $\mathcal{P}_\phi=(e_s/m_s)p_\phi$, being $e_s$ the charge and $m_s$ the mass of the considered species, and $p_\phi$ the canonical toroidal momentum, assuming an axisymmetric system. Moreover, $\mathcal{N},\alpha_w,T_w,\mathcal{P}_{\phi0},\Delta_{P_\phi},\lambda_0$ and $\Delta_\lambda$ are control parameters. In \cite{me}, the orbit theory has been described  through the  constant of motions (COMs) $\mathcal{P}_\phi,w,\lambda$,  where the canonical momentum $\mathcal{P}_\phi$, is treated as a spatial coordinate; the same choice is taken also here\footnote{At equilibrium the motion is  unpertubed and fields are stationary, so that it will be   considered only the guiding center transformation.}. \\  
 Together with (\ref{feq}), the regularized EDF,  
  \begin{equation}
 \label{feqRenorm}
f_{eq, R}=f_{eq} h_{eq},
\end{equation}
being 
\begin{equation}
\label{Renorm}
h_{eq}=\frac{\mathrm{H}(w_b-w) \delta_{\mathrm{confined}}}{1+(w_c/w)^{3/2}},
\end{equation}
has been proposed in \cite{me} as general plasma EDF for describing populations of particles encountered  in many tokamak scenarios. In (\ref{Renorm}),  $\mathrm{H}(w_b-w)$  is the \emph{Heaviside} step function which takes into account the presence of a mono-chromatic  source of \emph{birth} energy $w_b$. The factor $1+(w_c/w)^{3/2}$ mimics the \emph{Slowing-Down} behavior in energy, being $w_c$ the critical energy \cite{sivukhind,stix}, resulting from the relaxation of the considered species with bulk ions and electrons. The symbol  $\delta_{\mathrm{confined}}$ is the analytical condition for a particle whose orbit is mostly determined by $\mathcal{P}_\phi, w$ and $\lambda$, to be confined in the plasma volume\footnote{The explicit analytical expression of $\delta_{\mathrm{confined}}$ can be found in \cite{me} and will  not be reported here.}. 
This EDF has already been implemented in the hybrid code XHMGC \cite{XHMGC} and in the gyrokinetic code NEMORB \cite{NEMORB}. It has been shown that, by varying the EDF control parameters $\mathcal{N},\alpha_w,T_w,\mathcal{P}_{\phi0},\Delta_{P_\phi},\lambda_0$ and $\Delta_\lambda$,  (\ref{feqRenorm}) can represent anisotropic equilibria  as for the case of Neutral Beam Injection and Ion Cyclotron (or Electron Cyclotron) Resonance Heating. Moreover, it can also represent nearly isotropic equilibria as for the case of Slowing-Down \emph{alpha} particles and core thermal plasma populations.
In \cite{me} it has been proposed a heuristic derivation of $f_{eq}$ whilst, in the present work, a rigorous one is shown based on probabilistic principles and general hypothesis for deriving a class of EDFs which includes also the distribution function (\ref{feq}) and (\ref{feqRenorm}).\\
  The distribution function, $f_{eq}=f_{eq}(\mathcal{P}_\phi,w,\lambda)$, in (\ref{feq}) is an EDF because it depends solely on COMs. In this way the total time derivative is 
  \begin{equation}
  \dot f_{eq}=\dot{\mathcal{P}}_\phi \partial_{\mathcal{P}_\phi} f_{eq}+\dot{w} \partial_w f_{eq} + \dot{\lambda} \partial_\lambda f_{eq}=0,
  \end{equation}
  being $\dot{\mathcal{P}}_\phi=\dot{w}= \dot{\lambda}=0$.
  In an EDF, the dependency on COMs is commonly obtained by the transport \emph{Boltzmann} equation, and precisely by the \emph{kernel} of the \emph{Boltzmann} collision operator, $C_B$:
    \begin{equation}
  C_B(f_{eq})=0.
  \end{equation}
  Often the problem for solving the equilibrium of the \emph{Boltzmann} equation for plasmas is attacked in the following manner: \\
  1) give an analytical expression to the collision operator,\\
   2) find the distribution function that belongs to the kernel of the collision operator, \\
   3) check if a combination of such solutions is constant in time and\\
    4) try to express such combination of solutions as the function of a particular set of COMs.\\
  A simplification of the problem is realized just starting from the maxwellian distribution function which is the known solution of the \emph{Boltzmann} equation, expressed with the \emph{Landau} collision operator, for a gas of charged particles interacting via \emph{Coulomb} collisions.  
  The maxwellian EDF is further transformed into a \emph{local} maxwellian and, further, into a \emph{canonical} maxwellian to end on writing the EDF as a \emph{unbiased} canonical maxwellian \cite{angelino}. \\
  The problem is  that the maxwellian satisfies the \emph{Boltzmann} equilibrium for \emph{Coulomb}  interactions but it is not function of COMs, energy apart. Moreover, although the \emph{unbiased} canonical maxwellian is constant in time, it doesn't really belong to the standard \emph{Landau}  collision operator kernel. \\
  Here it is proposed an alternative way to reach an EDF, using the \emph{Bayes}' theorem;  the \emph{maxwellian} distribution function is considered  only as the prior probability distribution function (pdf), while the joint distribution function to have a certain probability to find a particle with given $\mathcal{P}_\phi$, $w$ and $\lambda$ are considered the final EDF expressed as the product of the conditional probability, to have such $\mathcal{P}_\phi$ and $\lambda$ once the energy of the particle is known, multiplied for the prior pdf. The obtained EDF, being different from the maxwellian EDF, cannot be anymore considered as the solution of the standard \emph{Landau}  collision operator but it will be shown that it belongs to the kernel of a \emph{Boltzmann} collision operator which is a modified \emph{Landau} collision operator.\\
   The problem of finding an EDF depending entirely and solely on COMs is addressed in section~2, while the form of the associated  \emph{Landau} collision operator is studied in section~3. 
 \\
 
\section{Probabilistic derivation of the equilibrium distribution function}

The position $x$ of a charged particle in a magnetic field is expressed by 
\begin{equation}
x=X+\rho,
\end{equation}
 where $X$ is the Guiding Center (GC) and $\rho$ is the vector \emph{Larmor} radius. Similarly  the  particle velocity $v$ is  
 \begin{equation}
 \label{v}
 v=\dot{x}=V+\sigma,
 \end{equation}
   where $V=\dot{X}$ is the GC velocity and $\sigma=\dot{\rho}$ is  the difference between the particle and the GC velocities. In the GC transformation,  the velocity is expressed by $v=v_\| b+v_\perp$, separating a component parallel to the magnetic field $B$ ($b$ is the unit vector of $B$) to the component perpendicular to it. At first it is possible to associate $V$ with $v_\| b$ and $\sigma$ with $v_\perp$, but this correspondence can be done only approximatively, being exactly only for constant and uniform magnetic field when the \emph{drift} velocity is zero. In general $V$ and $\sigma$ are not orthogonal vectors and the angle $a$ between them is important for the present derivation:
 \begin{equation}
 \cos a \equiv \frac{\sigma\cdot V}{|\sigma| |V|_s},
 \end{equation} 
 being $|V|_s=\mbox{sgn} (V\cdot B) |V|$, where the sign of $V$ depends on its orientation towards $B$. The angle $a$ belongs to $[0,\pi]$ if $V\nearrow B$, and $a\in[\pi,2\pi]$ if $V\searrow B$. $|V|_s$ generalizes the parallel velocity $v_\|$, and it is particularly useful when the GC transformation is taken at all perturbative orders as done in \cite{me3}.\\
  The value $\cos a$ is considered as the realization of the random variable $Cosa$. Such assumption can be also considered a consequence of the GC transformation which treats the gyro-angle $\gamma$ as an ignorable coordinate as if it is possible to randomly chose a value of $\gamma$ without affecting the equations of motion in GC coordinates. This can be done also for $\cos a$ because it depends strictly on $\gamma$.  The gyro-angle $\gamma$ can be assumed to be uniformly distributed between $[0,2\pi)$, whilst the distribution of $\cos a$ will depend on the considered physical scenario. Regardless of the scenario, it is assumed that such distribution is centered at zero.  A simple situation where the value of $\cos a$ is always \emph{zero}, is for  constant and uniform $B$, when $V$ is orthogonal to $\sigma$.\\
 It is here assumed that $Cosa$ is distributed as a Normal distribution function with a deviation depending on the parameter $k$. In this way, given all particles with same  GC position $X$, same value of the GC velocity  $|V|_s$ and  same kinetic energy $w$, then 
  \begin{equation}
  \label{cosa}
  Cosa \sim \mbox{N} (0, \kappa/2|V|),
  \end{equation} 
 where $\kappa$ will be determined later, at the end of the present section, to discriminate various tokamak scenarios.

From the knowledge of $w, |V|_s$ and $\cos a$, it is possible to compute $|\sigma|$ as below shown starting from the equivalence $2w=v^2=(V+\sigma)^2$. Indeed\footnote{The present derivation can also  be used on describing elementary particle decaying processes if the products are a pair of particles departing at right angles.},
 \begin{equation}
 \sigma^2 +2 |\sigma| |V|_s \cos a -(2w- V^2)=0,
 \end{equation}
 with the solution
\begin{equation}
|\sigma|=-|V|_s \cos a+\sqrt{V^2\cos^2 a +2w- V^2},
\end{equation}
rewritten for convenience as
\begin{equation}
\label{|sigma|}
|\sigma|=\sqrt{2w- V^2} \left[-\frac{|V|_s \cos a}{\sqrt{2w- V^2}}+\sqrt{\frac{V^2\cos^2 a}{2w- V^2} +1 }\right].
\end{equation}
As before, $|\sigma|$ can be considered the realization of a random variable because it is function of $\cos a$. The same can also be said for the following variable
\begin{equation}
m \equiv \frac{e_s\sigma^2}{2m_s \omega_c},
\end{equation}
where  $\omega_c$ is the \emph{cyclotron} frequency, defined as $\omega_c \equiv \dot \gamma$. The variable $m$ is an estimate of the magnetic moment $\mu$, here defined as: 
\begin{equation}
\label{mudef}
\mu \equiv \frac{w-V^2/2}{m_s\omega_c/e_s}.
\end{equation}
The choice on the above, unusual definition of the magnetic moment, comes from having considered ignorable the gyro-phase $\gamma$ and taken a null electric potential. In such case the single particle \emph{hamiltonian} is the kinetic energy expressed in the canonical variables: $(X,  P=V+e_s A/m_s)$ and $(e_s\gamma/m_s, \mu)$, $A$ being the vector potential, so that
\begin{equation}
\label{hamilt}
w=\frac{(P-e_sA/m_s)^2}{2}+\mu (m_s\omega_c/e_s),
\end{equation}  
from where the definition (\ref{mudef}) is taken. It is worth noticing that, from the above definition of the magnetic moment, $\mu$ is an exact COM. Indeed, the \emph{Hamilton}'s equations are: $(e_s/m_s)\dot{\gamma}=\partial_\mu w=\omega_c$ and $\dot{\mu}=(e_s/m_s)\partial_\gamma w=0$.\footnote{ When the electric potential, $\Phi(t,x)$, must be considered, it occurs to substitute the kinetic energy with the total particle energy: $w\to \varepsilon=w+(e_s/m_s)\Phi(t,x)=V^2/2+(e_s/m_s)\Phi(t,X)+\mu (m_s\omega_c/e_s)$, as described in \cite{me3}.} \\
The variable $m$ is a good estimate of $\mu$ when $\cos a \sim 0$. From equation (\ref{|sigma|}), the random variable $M$, of which the value $m$ is a realization, is explicitly written as
\begin{equation}
   M=\mu \left[\frac{\kappa Y}{2 \sqrt{2w- V^2}}+\sqrt{\left(\frac{\kappa Y}{2\sqrt{2w- V^2}}\right)^2 +1 }\right]^2,
   \end{equation}
   being
   \begin{equation}
   Y=-\frac{2|V|_s Cosa}{\kappa}.
   \end{equation}
   From (\ref{cosa}), the random variable $Y$ is distributed as $Y \sim  \mbox{N}(0,1)$.\\
    Introducing $\alpha$ and $\beta$:
   \begin{equation}
   \label{betalpha}
   \beta=\mu \  \mbox{ and } \ \alpha^2=\frac{\kappa^2}{2\mu (m_s\omega_c/e_s)}
   \end{equation}
    then $M$ is rewritten as
      \begin{equation}
   M=\beta \left[\frac{\alpha Y}{2}+\sqrt{\left(\frac{\alpha Y}{2}\right)^2 +1 }\right]^2,
   \end{equation}
   which is recognized as the \emph{Birnbaum-Saunders} (BS) \cite{BS} random variable: $M\sim \mbox{BS}(\alpha,\beta)$. The corresponding pdf, also known  as \emph{fatigue life} pdf, is
       \begin{equation}
 \label{BS}
 f_{BS}(m;\alpha,\beta)=\frac{1}{(2\pi\beta)^{1/2}}\frac{\beta+m}{2\alpha m^{3/2}}\exp  \left[ -\frac{1}{2\alpha^2}\left( \frac{m}{\beta}+\frac{\beta}{m}-2\right)\right],
\end{equation}
The BS pdf has been developed to model breakage due to cracks, when a material is subjected to cyclic stress and every $i$-th cycle leads to an increase in crack  extension of $Y=Y_i$. This crack grows under the repeated applications of a common cyclic stress pattern, until it reaches a critical size, when fatigue failure occurs. If the total extension of the crack is normally distributed, then $T$,  the time  until failure, is distributed as  a BS pdf (see \cite{kundu})\footnote{For plasmas, the analogy is interesting: the width of the crack is associated to the weighted  projection, depending on the scenario, of $V$ on $\sigma$. Such extension is zero for constant and uniform magnetic fields. 
 The random variable, $Y$, is assumed distributed as a \emph{Normal} pdf, and  one particle corresponds to one cycle as well as one crack extension, while the total extension corresponds to the sum  of $Y$ over the particles with assigned  $|V|_s$ and $w$. The probability density function of $m$, whose mean value is the magnetic moment, follows the same behavior of the time until failure of the stressed material.}.\\

The pdf of $m$ is a conditional one, because $\alpha$ and $\beta$ are supposed to be already known,  depending on $|V|_s$, $w$, $\kappa$ and $X$. A similar argument can be applied for the estimate $\bar{V}$ of $|V|_s$. Suppose to be able to make a measurement, affected by an error, of the GC velocity of a particle and repeat such measurement for all particles with same  $|V|_s$ and  same GC position $X$. The result $\bar{V}$ is assumed to follow a Gaussian behavior with a standard deviation $\Delta_V/\sqrt{2}$: 
\begin{equation}
\label{Vpdf}
 f_G(\bar{V};|V_s|,\Delta_V/\sqrt{2})=\frac{1}{\sqrt{\pi}\Delta_V}\exp\left[-\left(\frac{|V|_s-\bar{V}}{\Delta_V}\right)^2\right]
  \end{equation}
 The system is considered in the \emph{Maxwellian} state as prior pdf, expressed by the following exponential law:
\begin{equation}
\label{prior}
 g_{M}(w;T_w)= \frac{1}{T_w}\exp\left[-\frac{w}{T_w}\right].
\end{equation}
This assumption is  standard, but can also be generalized considering the $Gamma$ distribution:
   \begin{equation}
\label{prior2}
 g_{\Gamma}(w;\alpha_w,T_w)= \frac{(w/T_w)^{\alpha_w-1}}{T_w\Gamma(\alpha_w)} \exp\left[-\frac{w}{T_w}\right].
\end{equation}
   Some inferences are required to know how far the system is from being described by the prior distribution. The desired pdf is obtained from the conditional probability $\pi(\theta \mid x)$ to have $\theta$ once $x$ is given, where $\theta$ is the array of true values $\theta=(|V|_s,w,\mu)$ and $x$ the array of the ``estimated" quantities $x=(\bar{V},m)$. From \emph{Bayes}' theorem:
   \begin{equation}
   \pi(\theta \mid x)=\frac{\pi(\theta)f(x \mid \theta)}{\int_\Theta \pi(\theta^\prime) f(x \mid \theta^\prime) \, d\nu(\theta^\prime)},
   \end{equation}
 which means that the posterior probability, $\pi(\theta \mid x)$, is proportional to the product of the prior $\pi(\theta)$, (\ref{prior}) or (\ref{prior2}), multiplied for the conditional probability (\ref{BS}) and (\ref{Vpdf}):
  \begin{eqnarray}
\label{func}
\nonumber
\pi(\theta \mid m,\bar{V})&=& \frac{N(\mu+m)}{\pi\sqrt{2\mu} \ m^{3/2}\alpha T_w \Delta_V} \times \\
&&  \times \exp \left[ -\frac{(|V|_s-\bar{V})^2}{\Delta_V^2} \right] \exp\left[ -\frac{w}{T_w}\right]\exp  \left[ - \frac{(\mu-m)^2}{2\alpha^2 m\mu}\right].
\end{eqnarray}
The above distribution function is not a function of COMs, but it is possible to properly re-cast it into an explicit EDF form, firstly, by substituting the space parameters $\Theta \to \Theta_{COM}= (\mathcal{P}_\phi,w,\mu)$. This transformation can be realized from the linear dependency of the difference $\mathcal{P}_\phi-\psi$, and the magnitude of GC velocity divided for $\omega_c$:
\begin{equation}
\mathcal{P}_\phi-\psi=\mathcal{F}\frac{|V|_s}{\omega_c}.
\end{equation}
The function of proportionality, $\mathcal{F}$, can be computed explicitly from the GC transformation \cite{me3}. At lowest order, $\mathcal{F}$ is  $F$, defined in $B=\nabla \psi \times \nabla \phi + F \nabla \phi$. Indeed, at this order,  $|V|_s\to v_\|$ and $\mathcal{P}_\phi-\psi=Fv_\|/\omega_c$. The constant parameter $\mathcal{P}_{\phi 0}$ is conveniently defined to be
\begin{equation}
\mathcal{P}_{\phi 0}=\psi+\mathcal{F}\frac{\bar{V}}{\omega_c}.
\end{equation}
The pdf  becomes
 \begin{eqnarray}
\label{func}
\pi(\mathcal{P}_\phi,w,\mu \mid m,\bar{V}) &=& \frac{N(\mu+m)}{\pi\sqrt{2\mu} \ m^{3/2}\alpha T_w \Delta_V} \times \\
\nonumber
&& \times \exp \left[ -\frac{(\mathcal{P}_\phi-\mathcal{P}_{\phi 0})^2}{\Delta_{P_\phi}^2} \right] \exp\left[ -\frac{w}{T_w}\right]\exp  \left[ - \frac{(\mu-m)^2}{2\alpha^2 m\mu}\right].
\end{eqnarray}
The above distribution function represent an equilibrium only if also  $\alpha$ and $m$ are COMs. In terms of $\kappa^2$ and $m$ the following cases are considered.\\
As first choice, 
\begin{equation}
\label{first}
\kappa^2=\frac{ \Delta_\mu^2 (m_s/e_s) \omega_c}{m} \ \mbox{ and } \ m=s_0
\end{equation}
so that (\ref{func}) becomes
 \begin{eqnarray}
 \label{feq1}
f_{eq1}(\mathcal{P}_\phi,w,\mu)&=&N\frac{1+\mu/s_0}{\pi \Delta_\mu T_w \Delta_{P_\phi}} \times \\
\nonumber
&& \times \exp \left[ -\frac{(\mathcal{P}_\phi-\mathcal{P}_{\phi 0})^2}{\Delta_{P_\phi}^2} \right] \exp\left[ -\frac{w}{T_w}\right]\exp  \left[ - \frac{(\mu-s_0)^2}{\Delta_\mu^2}\right].
\end{eqnarray}
This case can be useful when the system is artificially prepared to select only particles with a magnetic moment $s_0$ or, better, for a classic gas of quantum charged particles with spin proportional to $s_0$. If the spread on the magnetic moment $\Delta_\mu$ is  sufficiently large, it can reasonably represent the bulk populations of plasma at thermal equilibrium. This simple distribution function seems to be original as not yet proposed in literature.\\
As second choice,
  \begin{equation}
  \label{second}
\kappa^2=\frac{ T_w\Delta_\lambda^2 (m_s/e_s)\omega_c w}{m} \ \mbox{ and } \ m=\lambda_0 w
\end{equation}
so that (\ref{feq1}) becomes
 \begin{eqnarray}
 \label{feq2}
f_{eq2}(\mathcal{P}_\phi,w,\mu)&=&\frac{N(\lambda_0+\mu/w)(w/T_w)}{\pi\lambda_0 \Delta_\lambda T_w \Delta_{P_\phi}w^{3/2}}\times \\
\nonumber
&& \times \exp \left[ -\frac{(\mathcal{P}_\phi-\mathcal{P}_{\phi0})^2}{\Delta_{P_\phi}^2} \right] \exp\left\{ -\frac{w}{T_w}\left[1+ \frac{(\mu/w-\lambda_0)^2}{\Delta_\lambda^2}\right]\right\}.
\end{eqnarray}
Substituting $\lambda=\mu/w$, the distribution function is almost identical with  (\ref{feq}) apart from the multiplicative factor  $(1+\lambda/\lambda_0)$, which is of minor importance in comparison to the exponential behavior. 
It is worth noticing respect to \cite{me} that here the derivation is probabilistic. Moreover, the constants of motion are exacts (in (\ref{feq}) they were computed only at leading order) and the power factor, $\alpha_w$ in (\ref{feq}) is a direct result: $\alpha_w=1$. Obviously, if another prior pdf is used, the final EDF will change; \emph{e.g.} if (\ref{prior2}) is used instead of (\ref{prior}), then the EDF becomes general as  (\ref{feq}), being $\alpha_w$  no longer fixed. It is worth noticing that if a \emph{Slowing Down} distribution function is used as the prior, being 
\begin{equation}
\label{SDdist}
f_{SD} (w)=\frac{\tau_S S_\alpha}{8\sqrt{2}\pi}\frac{\mathrm{H}(w_b-w)}{w^{3/2}+w_c^{3/2}},
\end{equation}
where $\tau_S$ is the \emph{Spitzer SD time} \cite{spitzer}, $w_b$ is the \emph{birth} energy, $w_c=v_c^2/2$ is the \emph{critical energy}, and $S_\alpha$ is a normalization constant, then, from (\ref{first}), the EDF becomes 
\begin{eqnarray}
 \label{feq3}
f_{eq3}(\mathcal{P}_\phi,w,\mu)&=&N\frac{\tau_S S_\alpha(1+\mu/s_0)}{8\sqrt{2}\pi^2 \Delta_\mu \Delta_{P_\phi}}\frac{\mathrm{H}(w_b-w)}{w^{3/2}+w_c^{3/2}} \times \\
\nonumber
&&\times \exp \left[ -\frac{(\mathcal{P}_\phi-\mathcal{P}_{\phi 0})^2}{\Delta_{P_\phi}^2} \right] \exp  \left[ - \frac{(\mu-s_0)^2}{\Delta_\mu^2}\right],
\end{eqnarray}
Otherwise, from (\ref{second}), it is obtained:
\begin{eqnarray}
 \label{feq4}
f_{eq4}(\mathcal{P}_\phi,w,\lambda)=&&N\frac{\tau_S S_\alpha(1+\lambda/\lambda_0)(w/T_w)^{-1/2}}{8\sqrt{2}\pi^2 \Delta_\lambda T_w^{3/2 } \Delta_{P_\phi}}\frac{\mathrm{H}(w_b-w)}{w^{3/2}+w_c^{3/2}} \times \nonumber \\
&& \times \exp \left[ -\frac{(\mathcal{P}_\phi-\mathcal{P}_{\phi0})^2}{\Delta_{P_\phi}^2} \right] \exp\left[ -\frac{w}{T_w}\left(\frac{\lambda-\lambda_0}{\Delta_\lambda}\right)^2\right],
\end{eqnarray}
already considered in \cite{me} for representing fast particles heated by Neutral Beam Injection.
\\
Other EDFs can be obtained by varying some assumptions made here concerning the random variable $Y$. Indeed, the BS is a particular pdf belonging to the family of \emph{Inverse Gaussian} pdfs \cite{life,jorgensen,seshandri} and it can be generalized as already studied in \cite{cordeiro11,bourguignon12}.\\
In the present derivation it is also furnished the method for consistently addressing the values of the control parameters. The parameters $\lambda_0$ and $\mathcal{P}_{\phi 0}$ are obtained from the estimate of the magnetic moment and the estimate of the magnitude of GC velocity, $m$ and $\bar{V}$, respectively, which characterized some particular orbits thus becoming representative of the considered population. If such quantities cannot be measured or estimated, it is not a problem because the functional form of the EDF is maintained. In this case, the EDF will be mainly used as a fitting model function. This doesn't mean that the parameters are devoid of any physical interpretation. It simply means that such parameters should be inferred from what is measurable as, for example, density, temperature or pressure of the considered plasma species might be. \\

\section{Associated Landau collision operator }

The \emph{Boltzmann} equation for a distribution function $f_s=f_s(t,x, v)$, in which $s$ indicates the species of particles with mass $m_s$ and charge $e_s$, is $\dot{f_s}=C_B(f_s)$, where $C_B$ is  the \emph{Boltzmann} collision operator. An important class of these operators consists of the following \emph{Fokker Planck}  operators:
\begin{equation}
C_{FP}(f_s)=\nabla_v \cdot (D_s \cdot \nabla_v  + d_s ) f_s, 
\end{equation}
being $D_s$ the \emph{diffusion} matrix and $d_s$ the collisional \emph{drag}. In such case, the \emph{Boltzmann} equation is said \emph{Fokker-Planck} equation and applies for describing soft collisions, i.e. binary collisions with only  little transfers of velocity changes to scattered  particles. If a background species of mass $m_{s^\prime}$ and charge $e_{s^\prime}$, is recognized to work as scatterers with a distribution function $f^\prime_{s^\prime}=f(t,x^\prime, v^\prime)$, then the \emph{scattering Fokker-Planck} operator would be  $C_{FP}(f_s)=\sum_{s^\prime} C_{FP}(f^\prime_{s^\prime},f_s)$ where the diffusion matrix and the collisional drag are functional of $f^\prime_{s^\prime}$: $D_s=D_s(f^\prime_{s^\prime})$ and $d_s=d_s(f^\prime_{s^\prime})$.
A smart representation of the above operator is given by the \emph{Landau} collision operator:
\begin{equation}
\label{LandauC}
C_L(f^\prime_{s^\prime},f_s)=\frac{\gamma_{s^\prime s}}{2} \nabla_v \cdot \int d^3 v^\prime \, U(u) \cdot \left( f^\prime_{s^\prime} \nabla_v f_s - f_s \nabla_{v^\prime} f^\prime_{s^\prime} \right).
\end{equation}
being $\gamma_{s^\prime s}$ a constant which is known for the \emph{Coulomb} collision case, and where
\begin{equation}
D_{s^\prime s}(f^\prime_{s^\prime})=\frac{\gamma_{s^\prime s}}{2} \int d^3 v^\prime \, U(u) f^\prime_{s^\prime},
\end{equation}
and
\begin{equation}
d_{s^\prime s}(f^\prime_{s^\prime})=\frac{\gamma_{s^\prime s}}{2} \int d^3 v^\prime \, U(u) \cdot \nabla_{v^\prime} f^\prime_{s^\prime}.
\end{equation}
The \emph{scattering} matrix $U$ is 
\begin{equation}
U=\frac{1}{|u|}\left( \mathbb{1} - \frac{u u}{u^2} \right),
\end{equation} 
for $u=u(v,v^\prime)$ that will be specified later.
A useful features of Landau collision operator is that it can be described by the \emph{RMJ} potentials \cite{RMJ}. Moreover, the main characteristic of $C_L$ is the assurance of the relaxation system to equilibrium thanks to the  \emph{entropy production}, generally defined by
\begin{equation}
\Theta \equiv -\int d^3 v C_B \log f_s \, ;
\end{equation} 
it can be shown that $\Theta \ge 0$ for any $f_s$ which satisfies the \emph{Boltzmann} equation written with the \emph{Landau} collision operator. Moreover, if \emph{e.g.} $C_L$ is adopted, the equilibrium, $f_{eq}$, is reached only if $C_{L}(f_{eq})=0$ when $\Theta=0$.

The standard application of  $C_L$ for deriving the EDF for a dilute plasma via \emph{Coulomb} interactions leads to the \emph{maxwellian} distribution function, $f_M$:
\begin{equation}
\label{maxwclassic} 
f_M=\frac{n}{(2 \pi T_w)^{3/2}}\exp \left(-\frac{w}{T_w} \right).
\end{equation}
Indeed, if  $f_M$ and $f^\prime_M$ are substituted in (\ref{LandauC}) for  same species and same $T_w$, then
\begin{equation}
\label{LandauMaxw}
C_L(f^\prime_M,f_M)=-\frac{\gamma_C}{2T_w} \nabla_v \cdot \int d^3 v^\prime \, U(v,v^\prime) \cdot (v-v^\prime) f_M f^\prime_M,
\end{equation}
being $\gamma_C$ constant for \emph{Coulomb} collisions.
Now,  if $u=v-v^\prime$ is the relative velocity between two colliding particles then $U\cdot (v-v^\prime)=0$ and also $C_L=0$. Such result led many people to the belief that the \emph{maxwellian} distribution function is the only equilibrium distribution function allowed for a plasma. 

Hence, a simple method to overcome this common belief and allows the EDF (\ref{feq1}) or (\ref{feq}) to represent a plasma in a magnetic field at equilibrium is now considered. The problem is resolved appropriately  changing the \emph{Landau} collision operator. The velocity vector $u$, taken as the relative velocity for the maxwellian equilibrium, changes to  $u_{p1}$, defined as
\begin{equation}
\label{uplasma}
u_{p1}=-\frac{T^\prime_w}{f^\prime_{eq1}f_{eq1}}\left( f^\prime_{eq1} \nabla_v f_{eq1} - f_{eq1} \nabla_{v^\prime} f^\prime_{eq1} \right),
\end{equation} 
and leaving the same $U$ matrix in the same $C_L$ operator.
From (\ref{uplasma}),
\begin{equation}
\label{uplasma1}
u_{p1}=v-v^\prime +\frac{e_sT_w}{m_s\Delta^2_\mu} \left[ \frac{\sigma}{\omega_c}\frac{2\left( \mu^2 -s_0^2\right)-\Delta_\mu^2}{\mu+s_0}-\frac{\sigma^\prime}{\omega^\prime_c}\frac{2\left( \mu^{\prime 2} -s_0^2\right)-\Delta_\mu^2}{\mu^\prime+s_0}\right],
\end{equation} 
at same temperature, $T^\prime_w=T_w$, same magnetic moment spread, $\Delta_\mu$, and same mean magnetic moment, $s_0$.
The same \emph{Landau} collision operator and the corresponding entropy production, bring the system to the EDF (\ref{feq1}), which is \emph{maxwellian}-like when $\Delta_\mu$  and $s_0$ are sufficiently large. The maxwellian limiting behavior is recognized also in (\ref{uplasma1}). Indeed, if $\Delta_\mu$ and $s$ are large, then $u_{p1}$ comes to be the same relative velocity which brings the system to the \emph{maxwellian} EDF. It is also worth noticing that the high $\omega_c$ frequency at the denominator in (\ref{uplasma1}) implies that the equilibrium is close to a \emph{maxwellian} also when the magnetic field is sufficiently strong or when the \emph{Larmor} radius, $\rho_L=|\sigma|/|\omega_c|$, is sufficiently small. Although such model doesn't accurately to describe the true interactions between charged particles, it can be suggested that the velocity vector $u_{p1}$ in (\ref{uplasma}) is the sum of the relative velocity, $v-v^\prime$, that takes into account \emph{Coulomb} collisions, and a new part representative of other interactions, as it can be the \emph{Ampere}'s interactions between currents (indeed, both $\sigma$  as $\sigma^\prime$ are the velocities of  charged particles moving on closed loops).  Moreover, if $\sigma \to 0$
then $u_p=V-V^\prime$, becomes the relative velocity of colliding  \emph{guiding particles},  that are particles with the same velocity and position of GC, and null \emph{Larmor} radius \cite{me3}. This means that the equilibrium distribution of guiding particles is, again,  maxwellian with energy $V^2/2$.
Same considerations are true if the EDF in (\ref{feq}) is chosen to represent the thermal population equilibrium with the only difference to compute $U$ for $u=u_{p}$:
\begin{eqnarray}
\label{uplasma2}
&&u_{p}=-\frac{T^\prime_w}{f^\prime_{eq}f_{eq}}\left( f^\prime_{eq} \nabla_v f_{eq} - f_{eq} \nabla_{v^\prime} f^\prime_{eq} \right)=\\
\nonumber
&&=v\left[ 1+\left(\frac{\lambda -\lambda_0}{\Delta_\lambda}\right)^2-\left(\alpha_w-\frac{3}{2} \right)\frac{T_w}{w} \right]+\sigma \frac{e_s}{m_s\omega_c \Delta_\lambda^2}\left[ \frac{2\left(\lambda^2-\lambda^2_0\right)-\Delta_\lambda^2T_w/w}{\lambda+\lambda_0}\right]+ \\
\nonumber
&&-v^\prime\left[ 1+\left(\frac{\lambda^\prime -\lambda_0}{\Delta_\lambda}\right)^2-\left(\alpha_w-\frac{3}{2} \right)\frac{T_w}{w^\prime} \right]-\sigma^\prime \frac{e_s}{m_s\omega^\prime_c \Delta_\lambda^2}\left[ \frac{2\left(\lambda^{\prime2}-\lambda^2_0\right)-\Delta_\lambda^2T_w/w^\prime}{\lambda^\prime+\lambda_0}\right],\\
\nonumber
\end{eqnarray} 
with same EDF's parameters, $T_w,\lambda_0,\Delta_\lambda$ and $\alpha_w$ (the latter parameter derives from the prior distribution in (\ref{prior2}) instead of the (\ref{prior})). The above expression is more complicated than (\ref{uplasma1}), which is the preferred choice, although in this context, it is not possible to exclude (\ref{uplasma2}). It is worth noticing that if $\Delta_\lambda \to \infty$ and $\alpha_w=3/2$, then 
\begin{equation} 
u_p \to v-v^\prime-(\sigma-\sigma^\prime) \frac{e_sT_w}{m_s w \omega_c(\lambda+\lambda_0)},
\end{equation}
and it becomes, again, easy to represent a maxwellian-like EDF, for $(m_s/e_s)\omega_c(\lambda+\lambda_0) \gg (T_w/w)$. 

Unfortunately, a choice is necessary because the same stratagem of modifying the \emph{Landau} collision operator cannot be applied twice for deriving two or more EDFs. However, there is a difference between the EDF in (\ref{feq1}) (or in (\ref{feq})) and the others, (\ref{feqRenorm}), (\ref{feq3}) and (\ref{feq4}). Indeed, the lasts are describing plasma populations in the presence of sources and losses (of particles or energies). The equilibrium distribution function $f_{eq, s} \in\{f_{eq,R},f_{eq 3}, f_{eq 4}\}$,  is obtained from the balance between the collisions with the background populations, \textit{i. e.} bulk ions and electrons represented by $f_b \in \{ f_{eq 1}$ (or $ f_{eq}\}$), and the source and the losses of particles and energies:
\begin{equation}
0=\dot f_{eq, s}=C_B(f_{eq, s})=\sum_b C_L(f^\prime_b,f_{eq, s})+ \mathrm{S}_s+\mathrm{L}_s.
\end{equation} 
The species labelled with $s$ can be \emph{alpha} products, $\mathrm{He}$ minority, energetic Deuterium, etc. 
In this way, the losses and the sources are modeled with the same parameters of the EDFs; indeed, 
\begin{equation}
\mathrm{S}_s+\mathrm{L}_s=-\sum_b\frac{\gamma_{b s}}{2} \nabla_v \cdot \int d^3 v^\prime \, U(u) \cdot \left( f^\prime_{b} \nabla_v f_{eq, s} - f_{eq, s} \nabla_{v^\prime} f^\prime_b \right),
\end{equation}
where the EDF of the bulk is $f^\prime_{b}=f^\prime_{eq 1}$(or $f^\prime_{b}=f^\prime_{eq}$), and $u=u_{p1}$(or $u=u_p$). This remark is quite important for tokamak plasmas because some of the sources, \emph{e.g.} radio-frequency antennas or neutral beam injectors, are controlled and can be appropriately modeled to be consistent with the desired bulk equilibrium. 
  
\section{Conclusions}

In the present work, the probabilistic derivation of axisymmetric plasma EDFs, making use of the \emph{Bayes}' theorem, has been explained. Four EDFs, (\ref{feq1}), (\ref{feq2}), (\ref{feq3}) and (\ref{feq4}), are obtained explicitly depending on if the mean magnetic moment is almost  constant for particles with different energies, or if it scales in magnitude with the particle energy, and if a \emph{Maxwellian} or a \emph{Slowing Down} is considered as prior pdf. The pdf  (\ref{feq1}) can represent a plasma in thermal equilibrium in a particular limit behavior, when the parameters $\Delta_\mu$ and $s_0$ are sufficiently large. The pdfs (\ref{feq2}) and (\ref{feq4}) are recognized to be the same EDF  already proposed in  Ref.\cite{me} which is known to be useful when external heating sources, \emph{i. e.} Ion Cyclotron Resonance Heating or Neutral Beam Injection, are employed \cite{me2} in experimental tokamak campaigns. The pdf (\ref{feq3}) represents a more realistic EDF for fusion products in axisymmetric tokamak. \\
The relevance of the present derivation resides in its generality since it only requires the  GC transformation of phase space coordinates and an axisymmetric ambient magnetic field. Moreover, such derivation doesn't depend on the detailed form of the axisymmetric magnetic field. This means that the derived functional form of the EDF is machine independent. Only the COMs $\mathcal{P}_\phi$ and $\mu$ are functions of  the equilibrium $B$ so that different values of the EDF parameters correspond to different scenarios. Respect to \cite{me}, where only the leading order approximation of  COMs in GC coordinates are used, in this case instead, the considered COMs, $\mathcal{P}_\phi, w$ and $\mu$ are all exact invariants. 
%Nevertheless, as it commonly happens when a \emph{Bayesian} inference argument is considered, those EDFs must be applied with care, always asking if  data are well fitted by such  probabilistic model, if the addressed parameters appear reasonable and how sensible the results are to the modeling assumptions.\\
In conclusion, the proposed EDFs is placed in a more ''thermodynamic" framework with a little change of the \emph{Landau} collision operator, which is still preserved in the form.

\ack 
This work was supported by Euratom Community
under the contract of Association between EURATOM/ENEA. It was also partly supported
by European Union Horizon 2020 research and innovation program under grant agreement
number 633053 as Enabling Research Project CfP-WP14-ER-01/ENEA Frascati-01.
The author would also like to thank S. Sportelli and C. Sportelli.


\section*{References}
\begin{thebibliography}{99}

\setlength{\itemsep}{-\parsep}

\bibitem{me}
{\sc C.~Di Troia}, 
\newblock {\em Plasma Physics and Controlled Fusion}, {\bf 54}, (2012) 105017.
\bibitem{XHMGC}
{\sc S.~Briguglio}, et al.,
\newblock {\em Phys. Plasmas}, {\bf 2}, (1995) 3711;
{\sc X.~Wang}, et al.,
\newblock {\em Physics of Plasmas}, {\bf 18}, (2011) 052504.
\bibitem{NEMORB}
{\sc A.~Bottino}, et al.,
\newblock{\em Plasma Physics and Controlled Fusion}, {\bf 53}, (2011) 124027.
\bibitem{angelino}
{\sc P.~Angelino}, et al.
\newblock{\em Physics of Plasmas}, {\bf 13}, (2006) 052517;
\bibitem{me3}
{\sc C.~Di Troia}
\newblock {\em  From charge motion in general magnetic fields to the non perturbative gyrokinetic equation}, \emph{submitted to} Physics of Plasmas. 
\bibitem{BS}
{\sc Z.W.~Birnbaum} and {\sc S.C.~Saunders},
\newblock {\em Journal of Applied Probability}, {\bf 6} 2, (1969) 319.
\bibitem{kundu}
{\sc D.~Kundu} 
\newblock \emph{Birnbaum-Saunders Distribution}, home.iitk.ac.in/$\sim$kundu/pala-BS-1.pdf
\bibitem{spitzer}
{\sc L.Jr~Spitzer} 
\newblock {\em Physics of Fully Ionized Gases}. Interscience, New York. 2nd Revised edition (1962), eq. 5-29
\bibitem{sivukhind}
{\sc V.~Sivukhind}
\newblock {\em Rev. Plasma Phys.}, {\bf 4}, (1966), 93.
\bibitem{stix}
{\sc T.H.~Stix}
\newblock {\em Plasma Physics}, {\bf 14}, (1972), 367.
\bibitem{life}
{\sc A.W.~Marshall} and {\sc I.~Olkin}
\newblock  {\em Life Distributions}, (Springer Series in Statistics 2007), chap. 13.
\bibitem{jorgensen}
{\sc B.~J{\o}rgensen}
\newblock {\em Statistical Properties of the Generalized Inverse Gaussian Distribution} (NewYork Berlin: Springer-Verlag 1982).
\bibitem{seshandri}
{\sc V.~Seshadri}
\newblock {\em The Inverse Gaussian Distribution} (Oxford Univ Press 1993).
\bibitem{cordeiro11}
{\sc G.M.~Cordeiro}, {\sc A.J.~Lemonte} and {\sc M.M.~Ortega},
\newblock {\em Statistics}, {\bf 47} 3, (2013) 626.
\bibitem{bourguignon12}
{\sc M.~Bourguignon}, {\sc R.B.~Silva} and {\sc G.M.~Cordeiro},
\newblock {\em Journal of Statistical Computation and Simulation}, {\bf 84} 12, (2014), 2619.
\bibitem{RMJ}
{\sc M.~N.~Rosembluth}, {\sc W.~M.~MacDonald}, and {\sc D.~L.~Judd},
\newblock {\em The Physical Review}, {\bf 107}, 1.
\bibitem{me2}
{\sc C.~Di Troia}, et al.,
\newblock {\em 24th IAEA Fusion Energy Conference; San Diego, CA (U.S.), 2012; IAEA-CN--197; TH/P6--21; ''Simulation of EPM Dynamics in FAST Plasmas Heated by ICRH and NNBI"}
\end{thebibliography}
  \end{document}